\RequirePackage{ifpdf}
\ifpdf 
\documentclass[pdftex]{sigma}
\else
\documentclass{sigma}
\fi

\begin{document}

\allowdisplaybreaks

\renewcommand{\PaperNumber}{057}

\FirstPageHeading

\renewcommand{\thefootnote}{$\star$}

\ShortArticleName{On the Applications of a New Technique to Solve Linear
DE, with and without Source}

\ArticleName{On the Applications of a New Technique\\ to Solve Linear
Dif\/ferential Equations,\\ with and without Source\footnote{This paper
is a contribution to the Vadim Kuznetsov Memorial Issue
`Integrable Systems and Related Topics'. The full collection is
available at
\href{http://www.emis.de/journals/SIGMA/kuznetsov.html}{http://www.emis.de/journals/SIGMA/kuznetsov.html}}}

\Author{N. GURAPPA~$^\dag$, Pankaj K. JHA~$^\ddag$ and Prasanta K. PANIGRAHI~$^\S$}

\AuthorNameForHeading{N.~Gurappa, P.K.~Jha and P.K.~Panigrahi}

\Address{$^\dag$~Saha Institute of Nuclear Physics, Bidhannagar, Kolkata 700 064, India}

\Address{$^\ddag$~Department of Physics, Texas A M University, TX 77843,
USA}

\Address{$^\S$~Physical Research Laboratory, Navrangpura, Ahmedabad 380 009, India}
\EmailD{\href{mailto:prasanta@prl.res.in}{prasanta@prl.res.in}}
\URLaddressD{\url{http://www.prl.res.in/~prasanta/}}

\ArticleDates{Received November 01, 2006, in f\/inal form March
29, 2007; Published online April 20, 2007}

\Abstract{A general method for solving linear dif\/ferential equations of
arbitrary order, is used to arrive at new representations for the
solutions of the known dif\/ferential equations, both without and with a
source term. A new quasi-solvable potential has also been constructed
taking recourse to the above method.}


\Classification{33C99; 81U15}

\section[A new procedure for solving linear differential equations]{A new 
procedure for solving linear dif\/ferential equations}

Linear dif\/ferential equations play a crucial role in various branches of
science and mathematics. Second order dif\/ferential equations routinely
manifest in the study of quantum mechanics, in connection with
Schr\"odinger equation. There are various techniques available to solve a
given dif\/ferential equation, e.g., power series method, Laplace
transforms, etc. Not many general methods applicable to dif\/ferential
equations of arbitrary order exist in the literature (see \cite{GA}  and references therein). We make
use of a general method for solving linear dif\/ferential equations of
arbitrary order to construct new representations for the solutions of the
known second order linear dif\/ferential equations, both without and with a
source term. This method has found applications in solving linear single
and multi variable dif\/ferential equations. The solutions of linear
dif\/ferential equation with a source and development of a new Quasi-Exactly
Soluble (QES) system are the new results of this paper.

\subsection[Case (i). Linear differential equation without
a source term]{Case (i). Linear dif\/ferential equation without
a source term}

After appropriate manipulation, any single variable linear dif\/ferential
equation can be brought to the following form
\begin{gather} \label{ie}
\left[F(D) + P(x,d/dx)\right] y(x) = 0,
\end{gather}
where $D \equiv x \frac{d}{dx}$, $F(D) = \sum_n a_n D^n $ is a diagonal
operator in the space of monomials spanned by $x^n$ and $a_n$'s are some
parameters. Here $P(x,d/dx)=\sum_{i,j} c_{i,j} x^{i} (\frac{d}{dx})^{j}$,
where $c_{i,j}=0$ if $i=j$. Notice that, since $F(D)$ is a diagonal
operator, $1/F(D)$ is also well def\/ined in the space of monomials. The
following ansatz
\begin{gather} \label{an}
y(x) = C_\lambda \left \{\sum_{m = 0}^{\infty} (-1)^m
\left[\frac{1}{F(D)}P(x,d/dx)\right]^m \right \} x^\lambda
\equiv C_\lambda \hat{G}_\lambda x^\lambda
\end{gather}
is a solution of the above equation, provided $F(D) x^\lambda = 0$ and the
coef\/f\/icient of $x^\lambda$ in $P(x,\frac{d}{dx})^{m} x^\lambda$ should be
zero \cite{heun}. In order to realise (\ref{an}) as a power series we
impose the requirement that $P(x,\frac{d}{dx})$ lowers (i.e.\ $c_{i,j}=0$
for $i\geq j $) or raises the  degree of monomials. Substituting equation~(\ref{an}), modulo $C_\lambda$, in equation (\ref{ie})
\begin{gather*}
\left(F(D) + P(x,d/dx)\right) \left\{\sum_{m = 0}^{\infty} (-1)^m
\left[\frac{1}{F(D)}P(x,d/dx)\right]^m \right \} x^\lambda  \nonumber\\
\qquad{}= F(D) \left[1 + \frac{1}{F(D)}P(x,d/dx)\right] \left \{\sum_{m =
0}^{\infty} (-1)^m \left[\frac{1}{F(D)}P(x,d/dx)\right]^m \right \}
x^\lambda \nonumber  \\
\qquad{}= F(D) \sum_{m = 0}^{\infty} (-1)^m
\left[\frac{1}{F(D)}P(x,d/dx)\right]^m  x^\lambda \\
\qquad {} + F(D) \sum_{m = 0}^{\infty}(-1)^m \left[\frac{1}{F(D)}P(x,d/dx)
\right ]^{m + 1} x^\lambda \nonumber \\
\qquad{}= F(D) x^\lambda - F(D) \sum_{m = 0}^{\infty} (-1)^m
\left[\frac{1}{F(D)}P(x,d/dx)\right]^{m + 1} x^\lambda \nonumber\\
\qquad{} + F(D) \sum_{m = 0}^{\infty}(-1)^m \left[\frac{1}{F(D)}P(x,d/dx)
\right ]^{m + 1} x^\lambda = 0 .
\end{gather*}

Equation (\ref{an}) connects the solution of a given dif\/ferential equation to
the monomials. In order to show that, this rather straightforward
procedure indeed yields non-trivial results, we explicitly work out a few
examples. Consider the Hermite dif\/ferential equation, which arises in the
context of quantum harmonic oscillator,
\begin{gather*}
\left[D - n - \frac{1}{2} \frac{d^2}{d x^2} \right] H_n(x) = 0.
\end{gather*}
Here $F(D) = D - n$ and $F(D) x^\lambda = 0$ yields $\lambda = n$. Hence
\begin{gather*}
H_n(x) = C_n \sum_{m = 0}^{\infty}(-1)^m \left[\frac{1}{D - n} (- 1/2)
(d^2/dx^2) \right ]^{m} x^n .
\end{gather*}
Using $[D,(d^2/dx^2)] = -2 (d^2/dx^2)$ it is easy to see that
\begin{gather*}
\left[\frac{1}{(D - n)} (- 1/2) (d^2/dx^2) \right ]^{m} x^n = (- 1/2)^m
(d^2/dx^2)^m \prod_{l=1}^m \frac{1}{(- 2 l)} x^n 
\end{gather*}
and
\begin{gather*}
H_n(x) = C_n \sum_{m = 0}^{\infty}(-1/4)^m \frac{1}{m!} (d^2/dx^2)^m
x^n
= C_n e^{- \frac{1}{4} \frac{d^2}{dx^2}} x^n,
\end{gather*}
this is a well-known result. Similar expression also holds for the Lagurre
polynomial which matches with the one found in~\cite{khare}. Below, we
list new representations for the solutions of some frequently encountered
dif\/ferential equations in various branches of physics and mathematics~\cite{gra}. 
Notice that, the cases of conf\/luent hypergeometric,
hypergeometric, Chebyshev type II and Jacobi solutions are given in~\cite{heun}, 
and reproduced here for the sake of completeness.

 \textit{Legendre polynomial}
\begin{gather*}
P_n(x) = C_n e^{- \left\{1/(2[D + n + 1]) \right\} (d^2/d x^2)}  x^n.
\end{gather*}

\textit{Associated Legendre polynomial}
\begin{gather*}
P_n^m (x) = C_n (1 - x^2)^{m/2} e^{- \left \{1/(2[D + n + m +
1])\right\}(d^2/d x^2)}  x^{n - m} .
\end{gather*}

 \textit{Bessel function}
\begin{gather*}
J_{\pm \nu}(x) = C_{\pm \nu} e^{- \{1/(2[D \pm \nu])\}x^2} x^{\pm \nu}.
\end{gather*}

\textit{Generalized Bessel function}
\begin{gather*}
u_{\pm}(x) = C_{\pm} e^{-\{ \beta \gamma^2 /(2 [D + \alpha \pm \beta \nu
])  \}x^{2 \beta}}  x^{\beta \nu - \alpha \pm \beta \nu} .
\end{gather*}

\textit{Gegenbauer polynomial}
\begin{gather*}
C_n^\lambda(x) = C_n e^{- \{1/(2[D + n + 2 \lambda ])\} (d^2/d x^2)}
x^n .\nonumber
\end{gather*}

\textit{Hypergeometric function}
\begin{gather*}
y_{\pm}(\alpha, \beta ; \gamma; x) = C_{\pm} e^{- \{(1/ (D +
\lambda_{\pm}) \}\hat{A}} x^{- \lambda_{\mp}},
\end{gather*}
where $\lambda_{\pm}$ is either $\alpha$ or $\beta$ and $\hat{A} \equiv x
\frac{d^2}{d x^2} + \gamma \frac{d}{d x}$. All the above series solutions
have descending powers of $x$. In order to get the series in the ascending
powers, one has to replace $x$ by $\frac{1}{x}$ in the original
dif\/ferential equation and generate the solutions via equation~(\ref{an}).
However, the number of solutions will remain the same. One can also
generate the series solutions by multiplying the original dif\/ferential
equations with $x^2$, and
then, rewriting $x^2 \frac{d^2}{d x^2} = D (D - 1) = F(D)$.

The solution for the following equation with periodic potential
\begin{gather} \label{pp}
\frac{d^2 y}{d x^2} + a \cos(x) y = 0
\end{gather}
can be found after multiplying equation (\ref{pp}) by $x^2$ and rewriting $x^2
\frac{d^2}{d x^2}$ as $(D - 1) D$ to be
\begin{gather*}
y(x) = \sum_{m , \{n_i\} = 0}^{\infty} \frac{(- a)^m}{m!} \left\{\prod_{i
= 1}^m \frac{(-1)^{n_i}}{(2n_i)!} \right\} \\
\phantom{y(x) =}{} \times   \left\{\prod_{r = 1}^m
\frac{\left(2 \left[m + \lambda/2 - r + \sum\limits_{i=1}^{m+1-r}n_i\right]\right)!} {\left(2 \left[m + \lambda/2
+ 1 - r + \sum\limits_{i=1}^{m+1-r}n_i\right]\right)!}
\right\} x^{2\big(m + \sum\limits_{i=1}^m n_i + \lambda/2\big)},
\end{gather*}
where $\lambda = 0$ or $1$. In the same manner, one can write down the
solutions for the Mathieu's equation as well.

\textit{Chebyshev polynomials $T_n(x)$ and $U_n(x)$}
\begin{gather*}
T_n(x) = C_n e^{- \left\{\frac{1}{2} \frac{1}{(D + n)}
\frac{d^2}{dx^2}\right\}}  x^n
 \end{gather*}
and
\begin{gather*}
 U_n(x) = C_n^\prime e^{- \left\{\frac{1}{2} \frac{1}{(D + n + 2)}
\frac{d^2}{dx^2}\right\}} x^n ,
\end{gather*}
 where $C_n$ and $C_n^\prime$ are appropriate normalization
 constants.

\textit{Jacobi polynomial}
\begin{gather*}
 J_n^{(\alpha,\beta)}(x) = \sum_{m=0}^\infty \left[\frac{1}{(D - n)
 (D + \alpha + \beta + 1)}
 \left(\frac{d^2}{dx^2} + (\beta - \alpha) \frac{d}{dx}\right) \right]^m x^n ,
 \end{gather*}
Schl\"af\/li, Whittaker and for that matter, any solution of a second order
linear dif\/ferential equation without a source term can be solved in a
manner identical to the above cases.

\subsection[Case (ii). Linear differential equation with a
source term]{Case (ii). Linear dif\/ferential equation with a
source term}

Consider,
\begin{gather*}
\left(F(D) + P(x,d/dx)\right) y(x) = Q(x).
\end{gather*}
Now the solution can be found in a straightforward way as
\begin{gather*}
F(D) \left(1 + \frac{1}{F(D)}P(x,d/dx)\right) y(x) = Q(x),
\end{gather*}
and
\begin{gather*}
y(x) = \sum_{m = 0}^\infty (-1)^m \left[\frac{1}{F(D)}P(x,d/dx)\right]^m
\frac{1}{F(D)} Q(x).
\end{gather*}
We list below a few examples for the above case.

\textit{Neumann's polynomial}
\begin{gather*}
O_n(x) = \left\{\sum_{r=0}^{\infty} (- 1)^r \left[ \frac{1}{[(D + 1)^2 -
n^2]} x^2 \right]^r \left(\frac{1}{[(D + 1)^2 - n^2]} \right) \right\}
  \\
\phantom{O_n(x) =}{}\times\big(x \cos^2 (n \pi/2)  + n \sin^2(n \pi/2) \big) .
\end{gather*}

\textit{Lommel function}
\begin{gather*}
w(x) = \sum_{m=0}^\infty (-1)^m \left[\frac{1}{(D^2 - \nu^2)}x^2\right]^m
\frac{1}{(D^2 - \nu^2)} x^{\mu + 1} .
\end{gather*}
 The cases of Struve, Anger and Weber functions are identical to the above ones.

It is a priori not transparent that all the solutions of a given
dif\/ferential equation can be obtained through the present approach. If $F$
is a polynomial of the same degree as the order of the dif\/ferential
equation, and has distinct roots, the linearly independent solutions can
be obtained through this approach. As has been pointed out in an earlier
paper \cite{Charan_JCAM} a given single variable dif\/ferential equation can
be cast in the desired form, as required by the present approach, in more
than one way through multiplication by powers of $x$. In a number of cases,
these lead to dif\/ferent solutions. However, the case of degenerate roots,
as also the case of inhomogeneous  equations needs separate consideration,
which we hope to investigate in future.

\section{A new quasi-exactly solvable model}

The above procedure is also applicable to dif\/ferential equations having
higher number of singularities e.g., Heun equation and its generalization
\cite{Guru_Heun}. The same can be used to generate quasi-solvable models.
It is worth mentioning a quasi-solvable model based on Heun's equation has
been studied; it has been shown that, this system lacks the $SL(2,R)$ of
many well-known QES model~\cite{TDLee,TDLee1}.  We now proceed to generalize this
system for obtaining a new QES potential. The dif\/ferential equation under
consideration is given by:
\begin{gather}\label{Heun}
f'' + \left\{ \frac{1}{2x} + \frac{1+2s}{x-1} +
\frac{1}{2(x+\epsilon^{2})}\right\}f' +  \frac{(\alpha\beta x- q
-\frac{\Omega}{x})}{x(x-1)(x+\epsilon^{2})}f = 0.
\end{gather}

When $\Omega=0$ the above reduces to Heun equation. Under the following
change of variable
\begin{gather*}
x = \frac {\sinh^{2}\frac{\rho y}{2}}{{1+
\frac{1}{\epsilon^{2}}+\sinh^{2}\frac{\rho y}{2}}},
\end{gather*}
 and with $\Psi = ( 1-x)^{s}f(x)$
 the above equation can be cast in the form of
Schr\"odinger eigenvalue problem. Here $s=(1-E/\rho^{2})^{1/2}$, where
$E$ is energy of the system. The constants $\alpha$, $\beta$ and $q$ in the
equation~(\ref{Heun}) are related to $s$ in the following form:
\begin{gather*}
\alpha=-\frac{5}{2}-s,\qquad \beta=\frac{3}{2}-s\qquad  \text{and}\qquad
q=(1-s^{2})(1+\epsilon^{2})-\frac{1}{2}s\epsilon^{2}-\frac{1}{4}(1-2\epsilon^{2}).
\end{gather*}

The corresponding potential is given by
\begin{gather*}
V(y) =\rho^{2}\left\{\frac{ 8 \sinh^{4}\frac{\rho y}{2}-
4(\frac{5}{\epsilon^{2}}-1)\sinh^{2}\frac{\rho y}{2}+
2(\frac{1}{\epsilon^{4}}-\frac{1}{\epsilon^{2}}-2)}{ 8 (1+
\frac{1}{\epsilon^{2}}+\sinh^{2}\frac{\rho y}{2})^{2}}+
\frac{\Omega}{\epsilon^{2}\sinh^{2}\frac{\rho y}{2}}\right\}.
\end{gather*}

For the purpose of f\/inding solutions we cast the generalized Heun equation
(\ref{Heun}) in the form, $[F(D)+P]y(x)=0$.     Multiplying by $x^{2}$, we
get
\begin{gather*}
\left[-4x^{2}\epsilon^{2}\frac{d^{2}}{dx^{2}}-2\epsilon^{2}x\frac{d}{dx}-4\Omega
\right]f(x) +\left[4x^{3}
(\epsilon^{2}-1)\frac{d^{2}}{dx^{2}}+2(3\epsilon^{2}-2+4s\epsilon^{2})x^{2}\frac{d}{dx}\right.\nonumber\\
\left.\qquad{}-4qx+4x^{4}\frac{d^{2}}{dx^{2}}+
8(1+s)x^{3}\frac{d}{dx}+4\alpha\beta x^{2}\right]f(x)=0.
\end{gather*}
In the above equation $ F(D)= -4\epsilon^{2}D^{2}+2\epsilon^{2}D-4\Omega$
and the condition $F(D)x^{\xi}=0$ yields
\[
\xi_{\pm}= \frac{1}{4}\left(1 \pm\sqrt{1-16\Omega/\epsilon^{2}}\right).
\]
 For polynomial solutions
either of the roots of the equation must be an integer. Taking $\xi_{-}=
m$ we obtain the following relation for the allowed values of $m$
\begin{gather*}
2m^{2}-m+2\Omega/\epsilon^{2}=0.
\end{gather*}

Let us consider a case in which $\Omega=-\epsilon^{2}/2$. Taking care of
normalization condition and positivity of energy, the allowed value of $m$
is $1$. This yields a solution of the form:
\[
 f(x)= a_{1}x + a_{2}x^{2}+ a_{3}x^{3}+\cdots.
 \]
  If the polynomial terminates at $x^{n-1}$ then the coef\/f\/icients
$a_{n}$ and $a_{n+1}$ should be zero. Using this constraint, we get the
following condition for $n$
\[
 s^{2}+(2n-1)+n^{2}-n-\frac{15}{4}=0,
 \]
giving the allowed values of $n$ as 2, 3. For $n=2$, $s=1/2$, $E =
\frac{3}{4}\rho^{2}$ and $\epsilon^{2}=1$ we get the wave function
\begin{gather*}
\psi(y)=N\left(\frac{2}{2+{\mathrm{sinh}}^{2}(\rho
y/2)}\right)^{1/2}\left( \frac{{\sinh^{2}(\rho y/{2})}}{{2+\sinh^{2}(\rho
y/2)}}\right),
\end{gather*}
where $N$ is the normalization constant. For $n=3$, $s=-\frac{1}{2}$,
which makes the wave function un-normalizable. This procedure may f\/ind
applicability to unravel the symmetry properties of Heun equation and its
generalization~\cite{Hooke}. We have earlier studied the symmetry
properties of conf\/luent Hypergeometric and Hypergeometric equations using
the present approach \cite{Charan_JCAM}. The fact that, the solutions are
connected with monomials makes the search for the symmetry rather
straightforward.

In the case when $\Omega \neq 0$, in equation (\ref{Heun}), the additional
term in Heun equation manifests in the potential in the Schr\"odinger
equation, as a term having $\frac{1}{x^2}$ type singularity at the origin.
This is very interesting since it is of the Calogero--Sutherland type, a
system much investigated in the literature. Like the Calogero--Sutherland
case, a Jastrow type factor in the wave function appears because of this
singular interaction. At a formal level, this interaction modif\/ies the
singularity at the origin without adding any new singularity. We note that
$x=0$ is still a~regular singular point, since the limit $x^{2}
\frac{\Omega}{x^{2}}$ as $x\rightarrow 0$ is f\/inite.

In conclusion, we have developed a new method to solve linear dif\/ferential
equations of arbitrary order, both without and with source term. Using
this method, we worked out a few examples for the case of second order
linear dif\/ferential equations. We obtained known, as well as, the new
representations for the corresponding solutions. The same approach is also
applicable to QES models based on Heun equation and its generalizations.
In particular a new potential of QES type is constructed through this
approach. We intend to analyze the symmetry properties of the Heun
equation through the present formalism.

\subsection*{Acknowledgements}

We thank C.~Sudheesh for
giving a careful reading to the manuscript and Vivek Vyas for many useful
comments.

\pdfbookmark[1]{References}{ref}
\LastPageEnding

\end{document}